\documentclass[prb,twocolumn,amsmath,amssymb]{revtex4}
\usepackage{graphicx}
\usepackage{dcolumn}
\usepackage{bm}

\begin{document}

\title{Excited two-dimensional magnetopolaron states in quantum well\\ of resonant tunnel junction}

\author{D.Yu.~Ivanov$^{1}$, M.V.~Chukalina$^{1}$, E.G.~Takhtamirov$^{2}$, Yu.V.~Dubrovskii$^{1}$,
 L.~Eaves$^{3}$, V.A.~Volkov$^{2}$,
 E.E.~Vdovin$^{1}$, J.-C.~Portal$^{4,6,7}$, D.K.~Maude$^{4}$, M.~Henini$^{3}$, G.~Hill$^{5}$.}
 \email{dubrovsk@ipmt-hpm.ac.ru}
\affiliation{%
$^{1}$ Institute of Microelectronics Technology RAS, 142432
Chernogolovka, Russia\\
$^{2}$ Institute of Radioengineering and Electronics RAS, Moscow,
Russia\\
$^{3}$ The School of Physics and Astronomy, University of
Nottingham, Nottingham NG7 2RD, UK\\
$^{4}$ Grenoble High Magnetic Field Laboratory, MPI-CNRS, BP166
38042 Grenoble Cedex 9, France\\
$^{5}$ Department of Electronic Engineering, University of
Sheffield, Sheffield S1 3JD, UK\\
$^{6}$ Institut Universitaire de France, 103, Boulevard
Saint-Michel, 75005 Paris, France\\
$^{7}$ INSA, F31077 Toulouse Cedex 4, France
}%

\date{\today}

\begin{abstract}
Tunnel spectroscopy is used to probe the electronic structure in
GaAs quantum well of resonant tunnel junction over wide range of
energies and magnetic fields normal to layers. Spin degenerated
high Landau levels ($N=2\div7$) are found to be drastically
renormalised near energies when the longitudinal optical-phonon
($\hbar\omega_{LO}$) and cyclotron energy ($\hbar\omega_{C}$) are
satisfied condition $\hbar\omega_{LO}=m\hbar\omega_{C}$, where
$m=1,2,3$. This renormalisation is attributed to formation of
resonant magnetopolarons, i.e. mixing of high index Landau levels
by strong interaction of electrons at Landau level states with
LO-phonons.
\end{abstract}

\maketitle

In polar semiconductors, such as GaAs, electrons interact with LO
phonons to form polarons, i.e. the bare electron states are
renormalized. If a magnetic field \textbf{B} is applied
perpendicular to the plane of the GaAs quantum well, 2D electron
states are quantized into Landau levels (LL) of index $N$. Filling
of empty Landau levels in the QW by tunnelling current causes
excitation of resonant magnetopolarons \cite{dsarma84, peeters85}
when resonant condition $n\hbar\omega_{LO}=m\hbar\omega_{C}$,
where $n,m$ are integers, is satisfied. The resulting
magnetopolarons can be measured experimentally by monitoring
density of states in the quantum well by tunnel spectroscopy.
Previously only interaction of ground Landau state $N=0$ with
$N=1$ and $N =2$ ones via LO-phonons have been detected by means
of phonon assisted tunnelling spectroscopy \cite{boebinger90}. It
was displayed as anticrossing of the position of the peaks related
to the double phonon assisted tunnelling into LL ($N=0$) and
single phonon assisted tunnelling into LL ($N =1,2$) in the fan
diagram.

In this work we present studies of electron structure of the QW in
a magnetic field normal to the well plane by means of tunnel
spectroscopy. The AlGaAs/GaAs/AlGaAs heterostructure was double
barrier structure incorporating a layer of InAs quantum dots (QD)
in the center of the well. The dots are charged and create a
considerable amount of disorder in the well. In this case the
number of elastic scattering assisted tunnelling events with only
energy conservation, but not momentum conservation is increased
considerably. It means that direct tunnelling between different
LL's in the emitter and collector is permitted and one can monitor
the density of states in a magnetic field in the QW by means of
both elastic and inelastic tunnelling spectroscopy. The phonon
assisted tunnelling channel is opened independently of sample
quality. As the result we have found strong interaction between
LL's of different indices ($N=2\div7$) in the well near energies
when the longitudinal optical-phonon ($\hbar\omega_{LO}$) and
cyclotron energy ($\hbar\omega_{C}$) are satisfied condition
$\hbar\omega_{LO}=m\hbar\omega_{C}$, where $m=1,2,3$. This was
attributed to formation of resonant magnetopolaron states, i.e.
mixing of different LLs by interaction with LO-phonons.
\begin{figure*}
\includegraphics{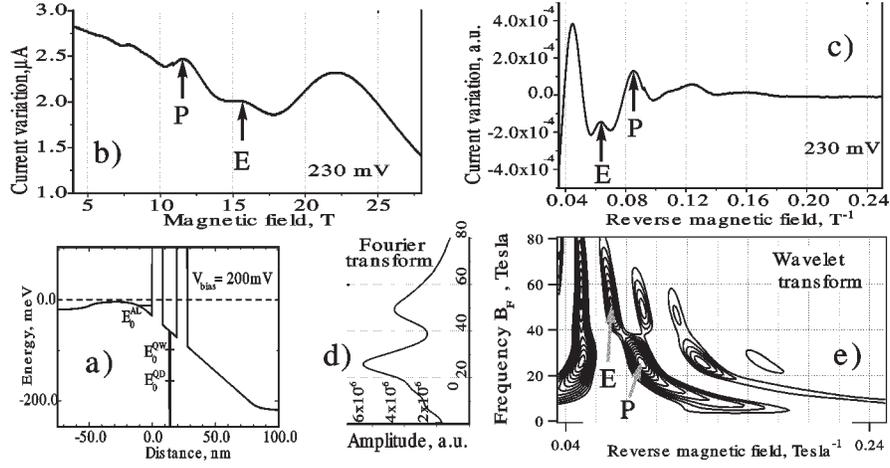}
\caption{\label{polaron1} (a) The band diagram of the structure
under voltage bias; (b) variation of tunnelling current versus
magnetic field at $V_{bias}=230$ mV; (c) variation of tunnelling
current versus reverse magnetic field at $V_{bias}=230$ mV; (d)
result of Fourier transform; (e) result of WL transform.}
\end{figure*}

Samples grown by MBE comprised (in the order of growth): a lightly
$Si$-doped, $300$-nm--thick GaAs layer ($N_{d} = 3\cdot10^{18}$
cm$^{-3}$); a $50.4$-nm--thick GaAs layer ($N_{d} = 2\cdot10^{17}$
cm$^{-3}$); a $50.4$-nm--thick undoped GaAs spacer layer; a
$8.3$-nm--thick $Al_{0.4}Ga_{0.6}As$ barrier layer; a $5.6$ nm
undoped GaAs layer; a $1.8$ monolayer (ML) InAs (with growth rate
$0.13$ ML/s to form InAs QD); a $5.6$-nm--thick undoped GaAs
layer; a $8.3$-nm--thick $Al_{0.4}Ga_{0.6}As$ barrier; a $50.4$ nm
undoped GaAs; a $50.4$-nm--thick GaAs layer ($N_{d} =
2\cdot10^{17}$ cm$^{-3}$); and a $300$-nm--thick GaAs layer
($N_{d} = 3\cdot10^{18}$) cap-layer. The samples were
characterised by photoluminescence (PL) spectroscopy, which
confirmed the existence of the corresponding QD or WL emission.
Ohmic contacts were obtained by successive deposition of
$AuGe/Ni/Au$ layers and subsequent annealing. Mesa structures,
with diameter between $20\ \mu m$ and $400\ \mu m$, were
fabricated by conventional chemical etching.

The band diagram of the structure under voltage bias is shown in
Figure~\ref{polaron1}(a). With the application of a sufficient
bias to the structure, an accumulation layer is formed adjacent to
the barrier, which serves as two-dimensional emitter. The details
of resonant tunnelling through this double barrier structures with
disorder introduced by QD was published earlier
\cite{dubrovskii01}. The tunnelling current oscillates with
varying magnetic field with constant bias applied to structure
(Fig.~\ref{polaron1}(b)). Peaks are observed, corresponding to
tunnelling into Landau states of the ground subband in the well
with and without phonon emission. We analyse only the data for
fields above $4 T$, when only the lowest energy Landau level is
occupied in the emitter. All measurements were carried out at
temperature $T=4$K.

Careful analysis of the experimental data at each $V_{bias}$
allowed us to identify all peaks in a magnetotunnelling spectrum.
Three different techniques have been used to do this:
identification of peak position in $1/B$ \emph{vs.} LL number,
Fourier analysis, and more sophisticated wavelet analysis. The
wavelet analysis \cite{dubrovskii01} is a mathematical tool, which
allows to decompose a signal in the locally confined waves
(Wavelets).
\begin{figure}
\includegraphics[width=0.5\textwidth]{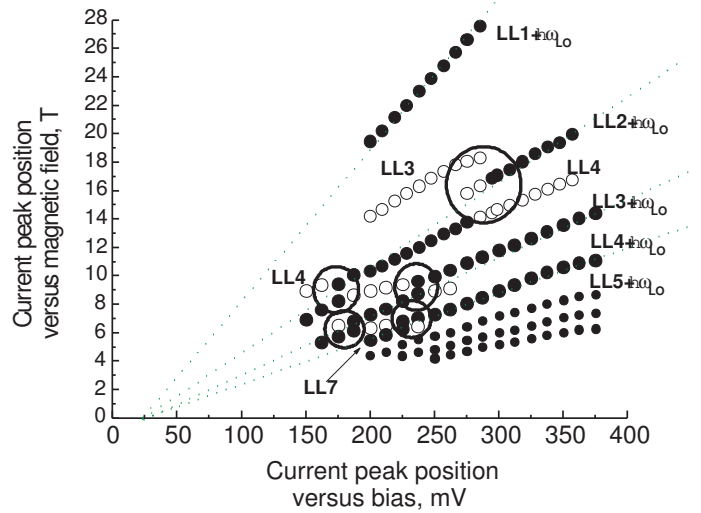}
\caption{\label{polaron2} Fan diagram summarised peak positions
versus magnetic field in tunnelling spectra. Solid circles -
position of peaks due to phonon assisted tunnelling. Open circles
- direct tunnelling from the ground Landau level $N=0$ in the
emitter to Landau level of higher indexes in the quantum well. In
the fan diagram it is easy to see regions where different levels
are anticrossing. These regions are indicated by circles.}
\end{figure}

The general expression for WL transform is $(Tf)(a,b)=\left|
a\right |^{-\frac{1}{2}}\int dtf(t)\psi(\frac{t-b}{a})$ , where
$\psi(t)$ is so called "mother" wavelet function. The transform
result is a function of two variables. Parameter $(\frac{1}{a})$
is analog of frequency in the Fourier transform. Each
$\psi(\frac{t-b}{a})$ is localized around $t=b$ . For WL
decomposition of our experimental data we have used the
Morlet~\cite{wlbook} "mother" function $\psi(t)=(\exp(i\gamma
t)-\exp(-\frac{\gamma^{2}}{2}))\exp(-\frac{t^{2}}{2})$ , which is
a trigonometric function modulated by Gaussian. The parameter
$\gamma$ determines the number of oscillations one wants to use
for the analysis and should be optimised in each specific case. In
more details the procedure of wavelet analysis is described in
paper \cite{web}.
\begin{figure*}
\includegraphics{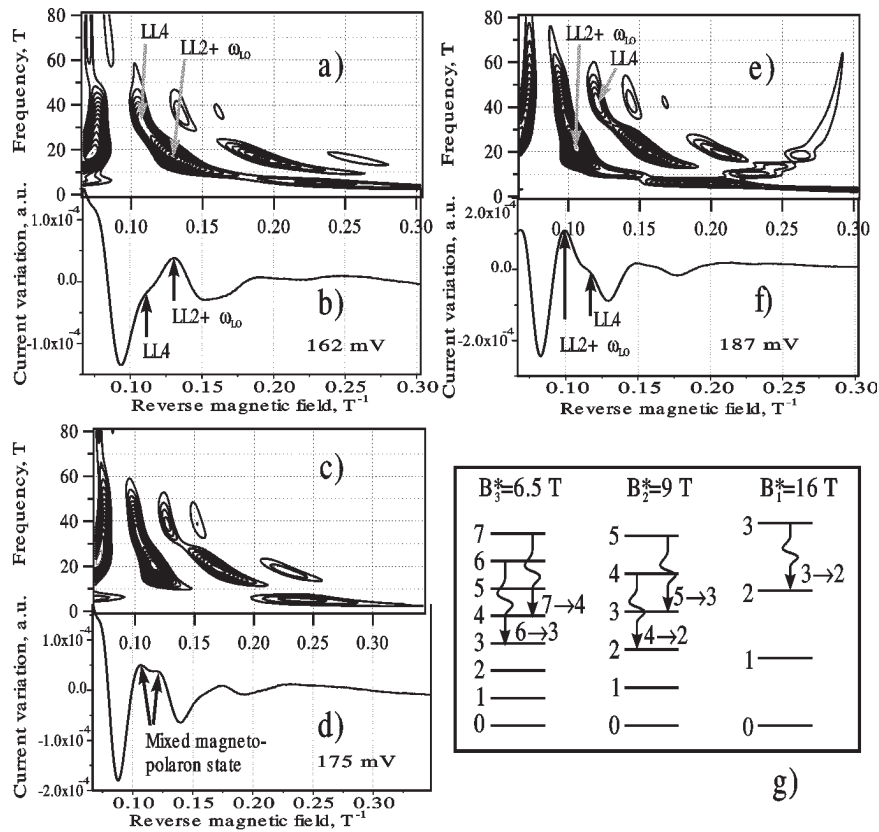}
\caption{\label{polaron3} This figure illustrates strong
interaction between N =2 and N=4 Landau levels in magnetic field
around $B=10$T when $2\hbar\omega_{C}=\hbar\omega_{LO}$, details
are in the text. (a) Wavelet transform of the tunnelling spectra
at $V_{bias}=162$ mV; (c) at $V_{bias}=175$ mV; (e) at
$V_{bias}=187$ mV; (b) variation of tunnelling current versus
reverse magnetic field for $V_{bias}=162$ mV; (d) for
$V_{bias}=175$ mV; (f) for $V_{bias}=187$ mV; (g) schematic
presentation of the identified magnetopolaron states.}
\end{figure*}

The result of WL analysis of tunnelling spectra recorded at $230$
mV is shown in Figure~\ref{polaron1}. Fig.~\ref{polaron1}(b) and
Fig.~\ref{polaron1}(c) show recorded signal versus magnetic field
and reverse magnetic field respectively. The result of Fourier
transform is shown in Fig~\ref{polaron1}(d). Figure 1(e) shows
result of WL transform. This is the 3-dimensional picture, where
the amplitude $A_{WL}=Re[(Tf)(\frac{1}{a},b) ]$ of the Wavelet
transform (z-axis) is presented versus frequency (y-axis) and the
inverse magnetic field (x-axis). As usual tunnelling with and
without phonon emission \cite{chukalina04} results in two sets of
oscillations periodical in reverse magnetic field in tunnelling
spectrum at constant bias, with frequencies
$B_{F}^{LO}=\frac{m^{*}}{e\hbar}(E_{Em}-E_{QW}-\hbar\omega_{LO})$
and $B_{F}^{El}=\frac{m^{*}}{e\hbar}(E_{Em}-E_{QW})$ respectively.
Here $m^{*}$ and $e$ are the effective mass and charge of an
electron; $\hbar$ is the Plank constant; $E_{Em}$, $E_{QW}$, are
the energies of the ground quasibound states in the emitter and in
the quantum well (QW) respectively, $\hbar\omega_{LO}$ is the
longitudinal optical phonon energy. Peaks in the tunnelling
spectrum related to phonon assisted tunnelling into LL belong to
low frequency set of oscillation and the elastic scattering peaks
to high frequency set. Fourier transform spectrum
(Fig.~\ref{polaron1}(d)) shows evidently two frequencies separated
by $\Delta B=21.4$ T. Than $\Delta B\frac{e\hbar}{m^{*}}=37$ meV,
which is equal to energy of LO-phonon ($36$ meV) in GaAs with good
accuracy as expected. Wavelet transform permits one to determine
the origin of each peak in the tunnelling spectrum. For example,
arrows labelled by "E" in Figures~\ref{polaron1}(b)
and~\ref{polaron1}(c) indicate the same peak. Appropriate maximum
in the amplitude $A_{WL}$ of the Wavelet transform
(Figure~\ref{polaron1}(e)), also indicated by arrow "E". Wavelet
data indicate that the peak belongs to the high frequency set of
oscillations and therefore is related to the elastic tunnelling
into LL with  $N=3$. The LL index could be easy determined by
standard procedure. In the same manner one can conclude that peak
labelled by "P" appears due to the phonon assisted tunnelling into
LL with $N=2$. (More precisely one should say that peak "P" is the
superposition of two peaks, one with higher amplitude related with
phonon assisted tunnelling and another of the smaller amplitude
with direct tunnelling into LL with $N=4$). In this way we have
identified all the peaks in the tunnelling spectra at different
bias voltages. Fan diagram in Figure~\ref{polaron2} summarised
peak positions versus magnetic field in tunnelling spectra. Solid
circles - position of peaks due to phonon assisted tunnelling.
Open circles - direct tunnelling from the ground Landau level
$N=0$ in the emitter to Landau level of higher indexes in the
quantum well.

Strong interaction between LL of different indexes have been
observed in tunnelling spectra $\hbar\omega_{LO}=m\hbar\omega_{C}$
with $m=1,2,3$. The details of this interaction between $N =2$ and
$N =4$ Landau levels ($m=2$) is shown in Figure~\ref{polaron3}, as
example. Tunnelling spectra versus reverse magnetic field for
three different biases are presented in Figures 3(b), (d), and
(f). Figures~\ref{polaron3}(a), (c), (e) show Wavelet transforms
of the tunnelling spectra, respectively. Since these peaks move
versus bias with different velocity, one could expect that without
interaction they should crossed at the intermediate bias voltage,
$V_{bias}=175$ mV in this case. Contrary we see typical
anticrossing behaviour of the peaks, first, the exchange of
oscillator intensity, second, the same peak is related to direct
tunnelling on one side or phonon assisted-tunnelling on another
side away from strong interaction point. Circles in the fan
diagram (Figure~\ref{polaron2}) indicate the regions where
different levels are anticrossing. Next magnetopolaron states have
been identified in this work:
\begin{eqnarray*}
  \hbar\omega_{LO}=\hbar\omega_{C}=E_{LL3}-E_{LL2} \\
  \hbar\omega_{LO}=2\hbar\omega_{C}=E_{LL5}-E_{LL3}=E_{LL4}-E_{LL2} \\
  \hbar\omega_{LO}=3\hbar\omega_{C}=E_{LL7}-E_{LL4}=E_{LL6}-E_{LL3}
\end{eqnarray*}
This is illustrated in Figure~\ref{polaron3}(g).

In conclusion, we have observed and identified a set of
two-dimensional resonant excited magnetopolaron states, i.e.
mixing of different high indexes Landau levels by optical phonons,
for electron tunnelling into a quantum well with embedded quantum
dots.

\begin{acknowledgments}
This work was supported by RFBR (04-02-16870, 02-02-22004), PICS
(1577), RAS programs "Quantum Macrophysics", and "Low-Dimensional
Quantum Nanostructures", FTNS program, EPSRC, and RS (UK). We are
grateful to Dr. H. Funke for the support during the software
development.
\end{acknowledgments}

\bibliography{polaron}

\end{document}